# ESTABLISHING SOFTWARE ENGINEERING DESIGN COMPETENCE WITH SOFT SKILLS

**LUIZ FERNANDO CAPRETZ**

WESTERN UNIVERSITY, LONDON, ONTARIO, CANADA
Email: lcapretz@uwo.ca

**Abstract** - For a long time, it has been recognized that the software industry has a demand for students who are well grounded in design competencies and who are ready to contribute to a project with little additional training. In response to the industry needs, an engineering design course has been developed for senior level students enrolled in the software engineering program in Canada. The goals of the course are to provide a realistic design experience, introduce students to industry culture, improve their time management skills, challenge them technically and intellectually, improve their communication skills, raise student level of professionalism, hone their soft skills, and raise awareness of human factors in software engineering. This work discusses the details of how this design course has been developed and delivered, and the learning outcomes that has been obtained.

## I. INTRODUCTION

There is more to producing software than just writing programs [1]. It is now widely recognized that the engineering of software systems has a pivotal role to play in the production of quality software systems that are produced on time, to budget, and to correct level of reliability. That puts software designers in the driver's seat of the high-tech revolution. Software is the secret elixir that transforms boring pieces off computer hardware into interactive tools capable of real magic.

The capstone course emphasizes the idea of what makes a good design as a key aspect within software engineering taking into consideration the most relevant soft skills that influence software development.

Software design is the study of the modern methods, technologies, languages, principles and practices that make it possible to conceive, create, validate and evolve complex software systems. The course emphasizes teamwork, and hones creative and entrepreneurial skills while putting the methods and techniques learned in past courses into real practice.

The project involves forming two or four-person software teams to analyze, design, build, test, and evaluate a software system to meet the requirements of a real independent user. Another objective of this course is the development of a new generation of innovators and entrepreneurs.

## II. RESEARCH QUESTION

Firstly, we address one fundamental question:

What happens during software development?

Table 1 shows the five stages of a software life cycle model and proposes an approach to conceptualize the points at which a particular dimension of personality trait may have some impact. A particular personality dimension influences each of the five phases to some extent [2], [3]. The theory behind each of the personality variables suggests that it is likely to impact some phases more than others. The personality dimension that appears most relevant to each phase is shown on the diagram and the rationale for each is described beneath. In other words, this diagram has been developed by overlapping some aspects of the MBTI theory and the tasks defining each stages of a software life cycle model, and shows the potential impact of MBTI dimensions on the software life cycle:

- Extroversion
- Introversion
- Sensing
- Intuitive
- Sensing
- Feeling
- Judging
- Perceiving

| Personality | Software Life Cycle Stages | | | | |
|---|---|---|---|---|---|
| Types | Analysis | Design | Coding | Testing | Maintenance |
| Extroversion | Yes | No | NoNoNo | | |
| Introversion | No | No | Yes | No | No |
| Sensing | No | No | Yes | YesYes | |
| Intuition | No | Yes | No | NoNo | |
| Thinking | No | Yes | Yes | No | No |
| Feeling | Yes | NoNoNoNo | | | |
| Judging | No | NoNo | Yes | No | |
| Perceiving | No | NoNoNoYes | | | |

**Table 1**
**Impact of Personality Typess on Software Live Cycle [4]**

## III. METHODOLOGY

The culminating design experience in software engineering is the capstone course, which that spans an entire academic year. In this course students use their acquired academic knowledge into practice and work on a project that deals with a realistic software engineering design. Students work in teams of four to tackle a comprehensive software design project, building on the overall undergraduate course materials previously offered through the software engineering program. Each team is supervised by a faculty advisor with whom they meet on a regular basis. An Agile software process (SCRUM) is followed to develop the software project.

Throughout the course, students are expected to achieve milestones related to the design process, including: problem definition and project proposal; generation and evaluation of concepts; engineering analysis and walkthrough, software prototype design, implementation and testing, software demos and presentation, and preparation of design documentation and project retrospective.

Deliverables during the year are a written project proposal, a project walkthrough, two sprints and design reviews, demos of releases one and two, an oral design presentation, and a project retrospective. The developed software systems and prototypes are judged by different stakeholders. Assessment of various course components are carried out the course coordinator with help from teaching assistants, the evaluation of the final demo is performed by the project advisor.

Once alternative solutions to the design problem have been conceived, those solutions need to be analyzed and then a decision must be taken on which solution is best suited for implementation. The students follow a systematic methodology in order to evaluate alternative design and assist in making a decision. The types of analysis include: functional analysis, soft skills, human factors (anthropometrics, ergonomics, etc.), product safety and liability, economic and market analysis, and legal factors (patents, standards and codes, contracts and agreements). The work is based on theory and general data collected in these work [5], [6], [7], [8], [9].

During the design process, students interact with one another and with the instructor to define the problem, specify the constraints and refine a solution and to communicate to others how to realize the solution. The communication patterns vary among groups within the design phases and include group presentations, group meetings, progress reviews, collaborative sites, and electronic communication.

The main forms of communication are:

1. Group meetings: Members of each group hold regular meetings to discuss issues related to their problem, to brainstorm and to delegate tasks. The members record group activities and outcomes, ideas, concepts, data, and other information.

2. Walkthrough: Members communicate geometric relationship using drawings, sketches, system architecture, and technology employed. They develop and practice good and effective attitudes towards giving and receiving constructive criticism and suggestions.

3. Progress reviews: The project leader supervises the progress in the different phases of the design problem. A check list of pending and completed milestones is created and updated during progress review presentations.

4. Electronic communication: Electronic mails, learning management system, collaborative tools, and webpages are used.

5. Software demonstration: There is no engineering of software without software, therefore the student must give a demo of their software, i.e., launch it and run while showing the main software functionalities.

6. Group presentations: At the end of the semester, each group will make an oral presentation using computer assisted presentation packages.

## IV. CONCLUSION

An environment has been created in which the students are participants in the entire design process, with the opportunity to experiment on the side, in which failure is considered part of the design process as well as part of the learning process. The most important aspects of the student experience in this design course is the recognition of the importance of team efforts and group dynamics, the practice of cooperation and appreciation of different talents, the exposure to the complete cycle of design from concept to final production in the form of prototype, and the integration of economic, legal, and human factors into the design decisions. The final presentations are watched and ranked by other teams in a contest for the best five projects. As the final presentations including final demos are pre-recorded videos publically posted on YouTube; students usually include their presentation URL in their CV/resumes.

★★★